\documentclass[10pt,conference]{IEEEtran}
\IEEEoverridecommandlockouts
\usepackage{cite}
\usepackage{amsmath,amssymb,amsfonts}
\usepackage{algorithmic}
\usepackage{graphicx}
\usepackage{textcomp}
\usepackage{xcolor}
\usepackage{tabularx}

\usepackage{tikz}
\usepackage{color}
\usepackage{pgfplots}
\usepackage{varwidth}

\usepgfplotslibrary{fillbetween}
\usetikzlibrary{patterns}
\usetikzlibrary{shapes, arrows}
\usetikzlibrary {arrows.meta}
\usetikzlibrary{calc,math}
\usetikzlibrary{positioning}
\usetikzlibrary{shapes.geometric,backgrounds}

\usepackage{graphicx,calc}
\usepackage{comment}

\usepackage{bm}

\begin{document}

\title{A Hybrid Quantum-Classical Approach to the Electric Mobility Problem }

\author{\IEEEauthorblockN{1\textsuperscript{st} Margarita Veshchezerova}
\IEEEauthorblockA{
\textit{Terra Quantum AG}\\
St. Gallen, Switzerland \\
mv@terraquantum.swiss} 
\\
\IEEEauthorblockN{5\textsuperscript{th} Sebastian Schmitt}
\IEEEauthorblockA{
\textit{Honda Research Institute Europe}\\ 
Offenbach am Main, Germany\\
Sebastian.Schmitt@honda-ri.de}
\and
\IEEEauthorblockN{2\textsuperscript{nd} Mikhail Somov}
\IEEEauthorblockA{
\textit{Terra Quantum AG}\\
St. Gallen, Switzerland \\
misom@terraquantum.swiss} 
\\ 
\IEEEauthorblockN{6\textsuperscript{th} Michael Perelshtein}
\IEEEauthorblockA{
\textit{Terra Quantum AG}\\
St. Gallen, Switzerland \\
mpe@terraquantum.swiss}
\and
\IEEEauthorblockN{3\textsuperscript{rd} David Bertsche}
\IEEEauthorblockA{
\textit{Terra Quantum AG}\\
St. Gallen, Switzerland \\
dbe@terraquantum.swiss} \\
\IEEEauthorblockN{7\textsuperscript{th} Ayush Joshi Tripathi}
\IEEEauthorblockA{
\textit{Terra Quantum AG}\\
St. Gallen, Switzerland \\
aj@terraquantum.swiss}
\and
\IEEEauthorblockN{4\textsuperscript{th} Steffen Limmer}
\IEEEauthorblockA{
\textit{Honda Research Institute Europe}\\ 
Offenbach am Main, Germany\\
Steffen.Limmer@honda-ri.de}
}

\maketitle

\begin{abstract}
We suggest a hybrid quantum-classical routine for the NP-hard \textit{Electric Vehicle Fleet Charging and Allocation Problem}. The original formulation is a Mixed Integer Linear Program with continuous variables and inequality constraints. To separate inequality constraints that are difficult for quantum routines we use a decomposition in master and pricing problems: the former targets the assignment of vehicles to reservations and the latter suggests vehicle exploitation plans that respect the battery state-of-charge constraints. 
The master problem is equivalent to the search for an optimal set partition. 
In our hybrid scheme, the master problem is reformulated in a quadratic unconstrained binary optimization problem which can be solved with quantum annealing on the DWave Advantage system. 
On large instances, we benchmark the performance of the decomposition technique with classical and quantum-inspired metaheuristics: simulated annealing, tabu search, and vector annealing by NEC.
The numerical results with purely classical solvers are comparable to the solutions from the traditional mixed integer linear programming approaches in terms of solution quality while being faster. In addition, it scales better to larger instances.
The major advantage of the proposed approach is that it enables quantum-based methods for this realistic problem with many inequality constraints.
We show this by initial studies on DWave hardware where optimal solutions can be found for small instances.


\end{abstract}


\section{Introduction}

Many industrial problems related to \textit{logistics}, \textit{planning}, \textit{scheduling}, or \textit{resource allocation} can be formulated as NP-hard optimization problems over discrete and continuous variables \cite{lp_applications}. Employing efficient algorithms may significantly reduce operational costs and increase profits - therefore, the search for computational advantage is crucial for competitiveness. In practice, operational problems are  usually solved with of-the-shell commercial solvers such as Gurobi; however, when the search for the exact solution is time-consuming we can use \textit{heuristic} algorithms to get good results in a reasonable time. 

The emergence of the \textit{quantum hardware} offers new opportunities for the design of efficient heuristics \cite{Preskill_NISQ}. Quantum heuristics leverage the laws of quantum mechanics to improve \textit{the approximation gap} or reduce the \textit{the time-to-solution}\cite{DWave_benchmark}. For instance, the \textit{tunneling effect} in the navigation of the energy landscape allows \textit{quantum annealing} \cite{QA_over_SA_advantage} to find optimal solutions under a certain condition \cite{Farhi_AA}. This condition -- namely the polynomially-bounded spectral gap -- is nevertheless difficult to guarantee, therefore, an experimental evaluation on difficult industrial problems is \textit{necessary} to decide on the practical potential of the heuristic.


Quantum algorithms adapted to near-future quantum hardware solve problems in so-called \textit{QUBO} (Quadratic Unconstrained Binary Optimization) formulations \cite{Preskill_NISQ}, i.e. all problem specifications are captured in a quadratic objective function over binary variables. At first glance, the NP-hard QUBO is a powerful model: traditional discrete problems \cite{Lucas_Ising} as well as some simplified industrial use-cases \cite{my_paper_1, tq_jsp} can be represented in QUBO without a huge resource overhead.

However, most real-world industrial problems exceed this simple framework: for instance, in the MILP (Mixed Integer Linear Program) model the solution space is typically restricted by many \textit{inequality and equality constraints} and variables are not necessarily binary. 
In theory, if all variable domains are discrete and bounded, the MILP can be reformulated as QUBO: discrete variables are encoded as binary string, \textit{slack variables} transform inequality constraints into equalities, and quadratic \textit{penalty terms} $M(a_j^Tx - b_j)^2$ (where $M$ is a large number) in the objective function enforce linear equality constraints $a_j^Tx = b_j$.

However, these obvious transformations lead to a large overhead in the number of variables in QUBO \cite{tq_jsp}. Thus, the size of the problems tractable on the near-future quantum hardware becomes strictly limited. In addition, 
penalty terms negatively impact the performance of quantum routines due to the additional energy scale separating feasible and infeasible solutions \cite{qubo_penality, Quantum_CG_master}. To address this obstacle, the works \cite{QA_modified_mixer, QA_modified_mixer_2} suggest to restrict the quantum evolution to the feasible subspace - but the protocol is difficult to put into practice. Alternately, a hybrid augmented Lagrangian method \cite{Kouki_2020} may perform well when the formulation has only a few constraints. 

We believe that a balanced interaction between quantum and classical routines is the most promising way to enable quantum enhancement for complex optimization problems \cite{tq_hybrid}. In this work, we introduce a \textit{hybrid approach} that delegates some operational constraints to a classical routine while leaving a difficult selection problem to the quantum heuristic.
\bigskip

We consider the problem of managing a fleet of electric vehicles (EV) previously considered in \cite{Honda_paper, Limmer2023}. In this problem, we search for an exploitation plan for a set of EVs over a discrete time horizon $T =\{0, \dots, t_{\max}\}$. We aim to fulfill as many reservations as possible with cars from our fleet. When cars are not used we can recharge them. The work \cite{Honda_paper} proves the NP-hardness of the problem and suggests a MILP formulation that is further optimized with the Gurobi solver\footnote{gurobi.com}. The solver fails to find optimal (or even good) solutions in one hour already for instances with 10 vehicles over a 48-hour time horizon - motivating the exploration of heuristic approaches.

In the original MILP from \cite{Honda_paper} a set of \textit{inequality constraints} ensure that the state-of-charge (SoC) of each EVs battery is always non-negative and doesn't exceed the battery capacity. These inequality constraints are particularly challenging for quantum routines. 
We decompose the problem into \textit{master} and \textit{pricing} problems: the master problem coordinates the collective solution while the pricing suggests new charging and utilization plans for individual EVs. 
The master problem is equivalent to the NP-hard Set Partition problem \cite{set_partition_np_hard} that can be naturally formulated as QUBO \cite{Set_packing_QUBO} -- following \cite{Quantum_CG_master} in our hybrid procedure we solve it with quantum annealing. The pricing problem deals with the charging schedule limited by the SoC-validity constraints. It uses a graph representation of possible individual actions at each time step, a path in the graph corresponding to a complete exploitation plan for one vehicle. By associating different weights to the edges of the graph in an iterative way we encourage the pricing problem to include or not a particular reservation in the plan.

We numerically evaluate our approach on realistic data\footnote{https://www.ac.tuwien.ac.at/research/problem-instances/\#evfcap}. As the actual quantum hardware can tackle the problems of relatively modest size, in addition to experiments on the \textit{DWave Advantage 6.1 system} for small instances (over 8-hour time horizon) we benchmark classical and quantum-inspired meta-heuristics on master problems for large instances (over 48-hour time horizon).

\paragraph*{Structure of the paper} In section \ref{section:PS} we introduce the problem, briefly recall the structure of the MILP formulation from \cite{Honda_paper}, and present the decomposition on the master and pricing problems. In section \ref{section:hybrid_algorithm} we present our hybrid approach and its potential applications. We report the results of our numerical experiments in section \ref{section:hybrid_algorithm} and discuss the insights in section \ref{section:discussion}.

\section{Problem statement}\label{section:PS}

In the Electric Vehicle Fleet Charge and Allocation Problem (EVFCAP) we consider a set $V$ of $n$ electric vehicles on a time horizon $T$ of $t_{\max}$ time steps each of duration $\Delta t$ (in our data $\Delta t = 15 \text{min}$). Each vehicle $v \in V$ has the same battery capacity $E^{cap}$ and an individual initial level of charge $E_{v, 0} \in [0, E^{cap}]$.
Reservations $r \in R$ ($|R| = r_{\max}$) have each a starting time $T_r^{start} < T$, an ending time $T_r^{end} \leq T$ and an expected energy consumption $E_r^{res} \in [0, E^{cap}]$. When vehicles are not used, they can be charged from the grid with power bounded by $p_{\max} \in \mathbb{R}^+$. 
The price of the grid energy varies in time and we denote its value at timestep $t$ with $c_t \in \mathbb{R}^+$. If a reservation is uncovered, i.e.\ not served by an EV from the fleet, it is fulfilled by a fuel car for the cost $c^{uncov} E_r^{res}$. 
Future costs are anticipated with a term $\alpha(E^{cap} - E_{v, t_{\max}})$ that penalizes low EV charging levels at the end of the time period. 
The target is to find a schedule of minimal cost satisfying all operational constraints.

\subsection{Compact formulation}

In \cite{Honda_paper} the problem is formulated as a Mixed Integer Linear Program (MILP). In what follows we refer to this MILP as \textit{compact formulation} (compact MILP). It has $(n + 1)r_{\max}$  binary variables: $x_{r, v}=1$ represent the assignment of the reservation $r$ to the vehicle $v$ and $y_r=1$ ($y_r=0$) indicates that a reservation is not assigned (assigned) to any vehicle. 
Continuous variables represent charging powers of vehicle $v$ at time $t$ as $p_{v, t} \in [0, p_{\max}]$. 
A multitude of constraints prevent conflicts, such as the assignment of two overlapping reservations to the same vehicle. In addition, for each vehicle inequality constraints ensure that the battery charging levels are within allowed bounds $[0, E^{cap}]$. 


\subsection{Extended formulation}

In this work, we focus on a slightly modified version of the original problem \cite{Honda_paper}: 
First, we restrict the charging powers to two discrete levels, $p_{n,t}\in\{0,p_{\max}\}$
(instead of the continuous values in the original formulation). 
Second, we discretize the energy levels $[0, E^{cap}] \rightarrow  \mathcal{E} =\{0, \dots, i\Delta E, \dots, i_{\max} \Delta E\} $ with equal spacing $\Delta E$ between these levels.
The initial SoC $E_{v, 0}$ and the  energies for reservations $E_r^{res}$ are rounded to the next lower and upper levels in $\mathcal{E}$, respectively.
Finally, we ignore "free" photovoltaic energy during the optimization process and simply integrate it into the final solution.

This modified problem formulation still captures all the relevant aspects and the complexity of the original problem setting.

\bigskip

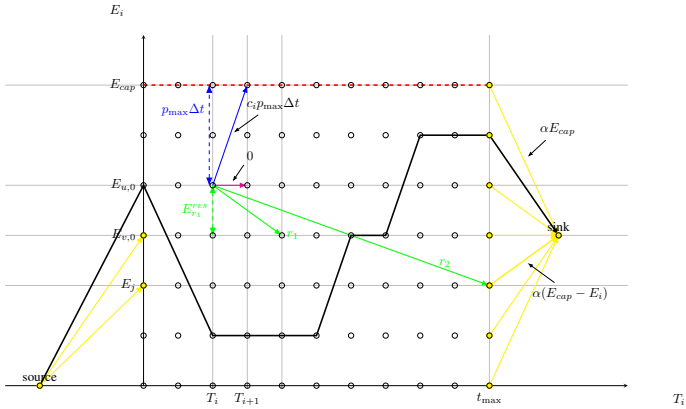
\begin{figure}[htbp]
    \centering
    \tikzset{>={Latex[length=2mm]}}
    \tikzset{pin arrow/.style={black, <-, >={Latex[length=1mm]}}}
    
    \begin{tikzpicture}[scale = 0.5]


      \begin{axis}[grid=both,
                    width=\textwidth, height=1.5*\axisdefaultheight,
                    ymin=0,ymax=7,xmax=14,xmin=-4,
                    xtick = {0, 2, 3, 4, 10},
                    xticklabels = {$0$,$T_i$,$T_{i+1}$, , $t_{\max}$},
                    ytick = {0, 2, 3, 4, 6},
                    yticklabels = {$0$, $E_{j}$, $E_{v,0}$, $E_{u, 0}$ ,$E_{cap}$},
                    minor tick num=1,
                    axis lines = middle, 
                    xlabel=$T_i$,ylabel=$E_i$,label style = {at={(ticklabel cs:1.1)}}]
      \addplot [domain=0:10, samples=10, color=red, very thick, dashed] {6};    
      
      \addplot [domain=2:3, samples=10, color=blue, thick, ->] {2*x} node[pos=0.5, pin={[pin edge={pin arrow}, black]60:$ c_{i} p_{\max} \Delta t$}] {};  
      
      \addplot [domain=2:3, samples=10, color=magenta, thick, ->] {4} node[pos=0.5, pin={[pin edge={pin arrow}, black]60:$0$}] {};
      
      \addplot [blue, dashed, <->] coordinates {(1.9, 4) (1.9, 6)} node [pos=0.75,anchor=east] {$p_{\max} \Delta t $};

      \addplot [domain=2:4, samples=10, color=green, thick, ->] {5-0.5*x} node [pos=1,anchor=west] {${r_1}$};
      
      \addplot [green, dashed, <->] coordinates {(2, 4) (2, 3)} node [pos=0.5,anchor=east] {$E_{r_1}^{res} $};
      
      \addplot [domain=2:10, samples=10, color=green, thick, ->] {4.5-0.25*x} node [pos=0.8,anchor=west, yshift=2pt] {$r_2$};
      
        \foreach \g in {0,1,...,5} {
            \addplot [yellow, mark = *, ->] coordinates {(10, \g) (12, 3)};   
        }

        \addplot [yellow, mark = *, ->] coordinates {(10, 6) (12, 3)} node[pos=0.5, pin={[pin edge={pin arrow}, black]60:$\alpha E_{cap}$}]{}; 

        \addplot [yellow, mark = *, ->] coordinates {(10, 2) (12, 3)} node[pos=0.5, pin={[pin edge={pin arrow}, black]280:$\alpha (E_{cap} - E_i)$}]{}; 
        
        \foreach \ezero in {2, 3} {
            \addplot [yellow, mark = *, ->] coordinates {(-3, 0) (0, \ezero)}; 
        }
        
        \addplot+ [nodes near coords, only marks,
        point meta=explicit symbolic, mark = o, mark size = 2pt]
        table [meta=label] {
        x y label
        -3 0 source
        12 3 sink
        };

        \foreach \g in {0,1,...,6} {
            \addplot+ [domain=0:10, samples=11, only marks, mark size=2pt, mark=o, draw=black] {\g};
         }

    \addplot [black, very thick, ->] coordinates {(-3, 0) (0, 4) (2, 1) (5, 1) (6, 3) (7, 3) (8, 5) (10, 5) (12, 3)};
    
    \end{axis}
    
    \end{tikzpicture}
    
    \caption{ \textbf{Graph of feasible exploitation scenarios.} For clarity we show arrows only for one internal node. 
    The graph contains one node per discrete SoC level per time step \textit{(black circles)} and two additional nodes called \textit{source} and \textit{sink}. Arrows correspond to different actions: charging \textit{(blue arrow)}, allocating to a reservation \textit{(green arrows)} and nothing \textit{(pink arrow)}. Only feasible actions are represented in the graph.
    We connect the source node to the initial SoC values $E_{v, 0}, v \in V$ (one arrow for each vehicle). All nodes corresponding to the final timestep $v_i = (i\Delta E, t_{\max})$ are connected to the sink node. 
    We assign a cost to each arrow \textit{(black subscript)}. Arrows corresponding to the charging at a timestep $T_i$ get the cost value of $c_i p_{\max} \Delta t$. The arrows going to the sink node have the cost $\alpha(E_{cap} - E_i)$. All other arrows have zero cost. \\
    The black path is an example of a feasible exploitation scenario that uses the vehicle $u$ and satisfies one reservation: in this scenario, we take the vehicle $v$ with initial charge $E_{v, 0}$, use it for one reservation starting at $t=0$, then charge it on some time intervals. 
    }
    \label{fig:exploitation_graph}
\end{figure}

 
A feasible exploitation scenario for one EV,  i.e.\ the  assignment to reservation(s) and the charging schedule, can be represented as a path on a \textit{weighted directed acyclic graph} $G = (\mathcal{V}, \mathcal{A})$ (see Fig.~\ref{fig:exploitation_graph}). Nodes $\mathcal{V}$ correspond to possible charge levels $\mathcal{E}$ at different time steps $\{0, \dots, T\}$; two auxiliary nodes \textit{source} and \textit{sink} help to encode the selection of a vehicle from the fleet and the value of the final SoC respectively. Arrows represent possible actions at different timesteps: selection of a particular vehicle, charging, allocation to a reservation, or nothing. 

The number of nodes depends on the discretization step $\Delta E$ and the number of timesteps $|T|$, while the number of arrows is linear on the size of the problem: we add one arrow from the source per vehicle and at most $|\mathcal{E}|$ arrows for each reservation. 

The EVFCAP problem can then be decomposed into two subproblems: 
i) the \textit{pricing problem} which suggest new promising exploitation scenarios, i.e. the paths in the graph ii) the \textit{master problem} which  selects one feasible exploitation scenario for each EV in the fleet from the subset of feasible exploitation scenarios. 
Each exploitation scenario comes with a cost, and the global objective is to minimize the total costs of selected scenarios plus the cost of unsatisfied reservations.

\subsection{Master problem}

We introduce one binary variable $\lambda_p\in\{0,1\}$ for each path $p \in \mathcal{P}$ from the source node to the sink node in the graph $G$. The cost of the path $c_p$ is a sum of the costs of all arrows in it: it corresponds to the sum of the grid energy costs (internal arrows) with the future costs (the arrow going to sink). In addition, as in the compact MILP, we take one variable $y_r \in \{0, 1\}$ per reservation $r \in R$; $y_r = 1$ implies that the reservation $r$ is unsatisfied.

The master problem may be formulated as Set Partition:

\begin{align}
    \min & \sum_{p \in \mathcal{P}} c_p \lambda_p + c^{uncov}\sum_{r \in R} E_r^{res} y_r & \label{eq:extended_objective}\\
     & \sum_{ \substack{ p \in \mathcal{P}: \\ r \in p} } \lambda_p + y_r = 1, & \forall r \in R \label{eq:reservation_const}\\
     & \sum_{ \substack{ p \in \mathcal{P}: \\ v \in p} } \lambda_p = 1, & \forall v \in V \label{eq:vehicle_const} \\
     & \lambda_p \in \{0, 1\}, & \forall p \in \mathcal{P} \label{eq:lambda_are_binary}
\end{align}

where
\begin{itemize}
    \item $r \in p$ in (\ref{eq:reservation_const}) means that for some $E_j$ the arrow $a_r: (E_j, T_r^{start}) \rightarrow (E_j - E_r^{res}, T_r^{end})$ corresponding to the reservation $r$ is in the path $p$. The constraint (\ref{eq:reservation_const}) implies that the reservation can be satisfied at most once.
    \item  In (\ref{eq:vehicle_const}) the notation $v \in p$ means that the arrow taken from the source node corresponds to the vehicle $v$. The constraint (\ref{eq:vehicle_const}) means that we have to select \textbf{precisely} one path for each vehicle (it can be trivial).
\end{itemize} 

The Set Partition problem (alternately called Exact Set Cover) is NP-hard \cite{set_partition_np_hard} as is its approximation within the factor $\ln n$ \cite{sp_approximation}. Typical instances issued from the EVFCAP data are difficult for the of-the-shell solvers: for example, the Gurobi solver does not find optimal solutions within one hour for $12$ out of $30$ instances with $t_{\max} = 192$, $n=20$ and $r_{\max} = 320$ from our dataset. 
On the other side, no inequality constraints are present in this formulation, which allows for an efficient mapping to a QUBO form. 
Therefore, if the quantum annealing for the QUBO formulation rapidly returns good-quality solutions, it can improve both the \textit{gap} and the \textit{runtime} of the proposed decomposition scheme.

\subsection{Pricing problem}

The number of paths $|\mathcal{P}|$ in the graph $G$ is exponential in the size of our instance. Therefore, we can't directly solve even \textit{the relaxed version} (where $\lambda_p \in [0, 1]$). We use so-called \textbf{column generation} \cite{CG_vanderbeck_PhD} originally introduced in \cite{CG_cutting_stock} to circumvent this obstacle. In column generation, while solving the relaxation we do not consider all variables $\mathcal{P}$ at once, but rather a restricted subset $\mathcal{P}' \subset \mathcal{P}$. We add new variables to $\mathcal{P}'$ \textbf{only if} they can improve the solution of the relaxation. 

Promising variables are found in the \textit{pricing routine} that searches for violated cuts in the dual problem. 
Indeed, each variable in the primal problem (whether in $\mathcal{P}'$ or not) corresponds to a constraint in the dual problem, and by the duality theorem if the dual solution is feasible, then the primal solution is optimal \cite{strong_duality}. 

In our case, the pricing is equivalent to the search of the \textit{shortest path} between \textit{source} and \textit{sink} nodes where edges corresponding to vehicle selection and reservations get costs determined by the dual solution of the restricted problem.

\section{Hybrid quantum-classical approach} \label{section:hybrid_algorithm}

\tikzset{
  scope/.style={execute at end picture={
        \begin{scope}[on background layer]
          \draw[black!15,fill=black!5,rounded corners=1ex] (current bounding box.south west)
                    rectangle (current bounding box.north east);
          \node[draw,fill=green!10,rectangle,anchor=west,inner sep=1pt,minimum width=4ex, xshift = 3cm, yshift = 0.1cm] at (current bounding box.north){#1};
        \end{scope}
    }},
}
\tikzset{
    max width/.style args={#1}{
        execute at begin node={\begin{varwidth}{#1}},
        execute at end node={\end{varwidth}}
    }
}
\tikzstyle{my style}=[draw, rectangle, rounded corners, align=center, text centered]
\tikzstyle{connector}=[draw, thick, -latex']
\begin{figure*}[htbp]
\begin{tikzpicture}[node distance = 1cm]
\node[fill=green!15, max width = 2.1cm, my style] at (0, 0) (mp){Master Problem \\ (variables $\mathcal{P}$)};
\node[below right= 0.5cm and -2.6cm of mp](relaxation){
\begin{tikzpicture}[scope={Solving the relaxation}, node distance = 1cm]
        \node[my style, max width = 2.1cm] (rmp) at (0, 0) {Restricted \\ Master \\ Problem \\ (variables $\mathcal{P}'$)};
        \node[my style, max width = 1.7cm, right = of rmp] (rrmp) {Relaxation \\ ($\lambda_p \in [0, 1]$)};
        \node[my style, max width = 1.5cm, right = of rrmp] (duals) {Get \\ dual \\ values};
        \node[my style, max width = 1.5cm, right = of duals] (pricing) {Solve \\ Pricing \\ problem};
        \node[trapezium, draw, text centered, trapezium left angle=60, trapezium right angle=120, minimum height=2em, align = center, right = of pricing, fill=blue!10] (improving) {$p$ improves \\ the solution? };
        \node[below = of improving] (yes){\textbf{Yes}};
        \node[right = of improving] (no) {\textbf{No}};
        \path[connector] (improving) -- (yes);
        \draw[connector] (yes) -| (rmp) node [midway, xshift = 5cm]{$\mathcal{P}'\rightarrow \mathcal{P}' \cup \{p\}$};
        \path[connector] (rmp) -- (rrmp);
        \path[connector] (rrmp) -- (duals) node[midway, above]{Solve};
        \path[connector] (duals) -- (pricing);
        \path[connector] (pricing) -- (improving) node[midway, above]{$p$};
        \path[connector] (improving) -- (no);
\end{tikzpicture}
};
\draw[connector](mp.south) -- ++ (0, -0.9cm); 
\node[my style, right = of relaxation, yshift=0.5cm](rsolved){\textbf{Relaxation} \\ \textbf{solved}};
\draw[connector] (rsolved) ++ (-2cm, 0) -- (rsolved) node[midway, above] {$\mathcal{P}'$};
\node[draw, rectangle, dashed, right = of mp, xshift = 5cm] (branching) {Branching};
\draw[dashed, thick, -latex'](rsolved) |- (branching) node [midway, above, xshift=-2cm] {Exact Branch \& Price}; 
\draw[dashed, thick, -latex'] (branching) -- (mp);
\node[my style, below = of rsolved, fill=red!20] (heur) {Solve \\ \textbf{Integer} \\ Master Problem \\ over variables in $\mathcal{P}'$  };
\draw[connector](rsolved) -- (heur) node [midway, left] {Heuristic};
\end{tikzpicture}
\caption{The Hybrid quantum-classical algorithm for the EVFCAP problem. The red node (solution of the integer master problem) is delegated to the quantum annealing. Dashed part shows the workflow of the traditional Branch \& Price.}
\label{fig:algorithm}
\end{figure*}
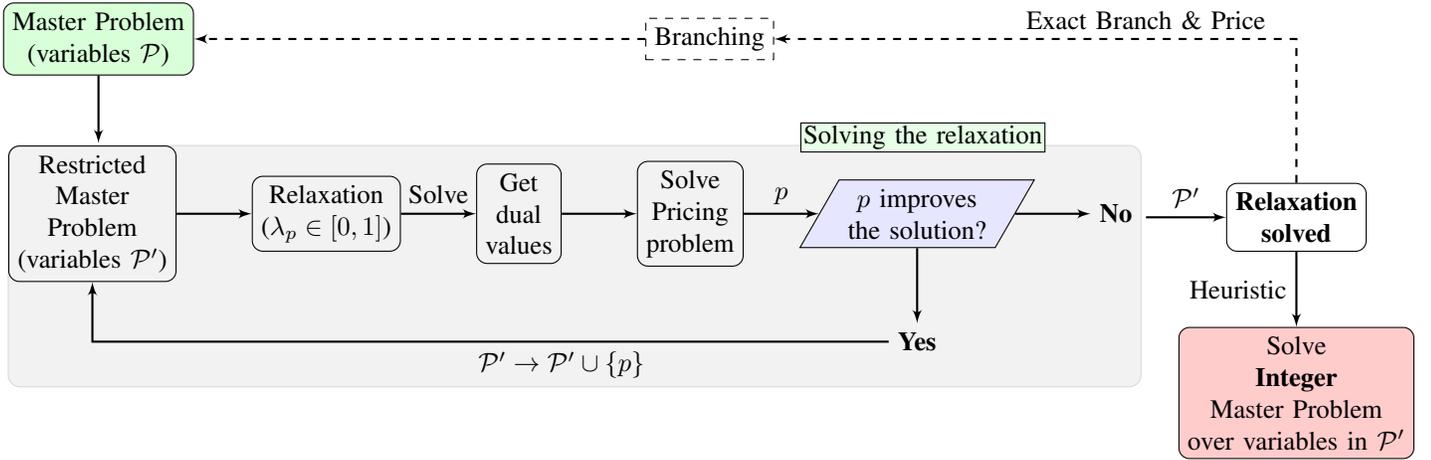

We suggest a hybrid approach that uses a classical column generation technique to build an instance of the Set Partition problem, that is further transformed into a QUBO form and solved with quantum heuristics (see Figure \ref{fig:algorithm}). 

In the first (classical) part we solve the \textit{relaxed version} of the master problem: we start from a restricted set of variables that guarantee the existence of a feasible solution (trivial plans) and iteratively add variables that may improve the relaxed solution. If the pricing problem fails to find a "promising" variable, the relaxation is solved to optimality. In such case we take all generated variables ($\mathcal{P}'$) and consider an \textit{integer} master problem over them - this ILP is equivalent to a search of an optimal set partition. 

 In the traditional Branch \& Price \cite{CG_vanderbeck_PhD}, a fractional solution for relaxation leads to the branching, and the variables may be regenerated in every node of the branching tree. The regeneration is necessary to find an exact optimum as an optimal \textit{integer} solution may involve variables that don't appear in the solution process for the \textit{relaxed} linear program. The quantum-assisted procedure presented in \cite{Quantum_CG_master} integrates the quantum solver as a \textit{primal heuristic} in the traditional Branch \& Price scheme.

 In contrast, in our approach, the variables are generated only once in the so-called root master problem. Compared to Branch \& Price we significantly reduce the running time at the cost of the optimality guarantees. In a nutshell, we obtain a heuristic method, where the column generation presents candidate exploitation plans for individual vehicles that are further combined in the master problem. 

Our hybrid approach can be applied in the same contexts as the Branch \& Price (or heuristic Branch \& Price) when, in addition, the time-to-solution is an important performance metric. We recall that the Branch \& Price is particularly suitable for complex planning and logistics problems where difficult (nonlinear) constraints restrict the set of possible solutions \cite{CG_book}.

\section{Numerical results}\label{section:numerical_results}

\begin{figure*}[htbp]
\centerline{\includegraphics[width=\textwidth]{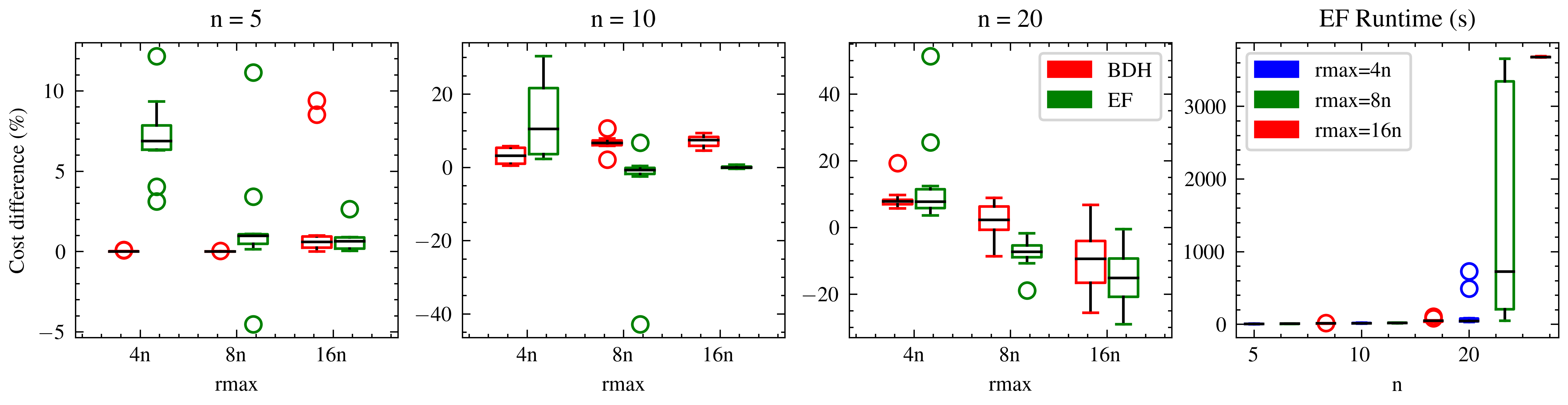}}

    \caption{Solution quality and runtime (in seconds) on instances for various numbers $n$ of EVs, maximal number of reservations $r_{\max}$ and for $t_{\max} =192$. Relative cost changes are given with respect to the 1-hour compact MILP solution.
    We report the values for the Bender-decomposition-based heuristic from \cite{Honda_paper} (BDH) and for the approach proposed here (EF) on instances with $t_{\max} = 192$.
    \\
    The first three plots demonstrate the relative difference in the cost value obtained by both heuristics. We observe that on most difficult instances (with $n=20$ or with $n=10$ and $r_{\max} > 8n$) the solution quality of the EF heuristic is systematically higher compared to BDH. For the largest instances ($n=20$), the negative relative cost (with respect to the MILP solution) demonstrates the advantage of the heuristics over the time-limited Gurobi solution.
    \\ 
    The rightmost plot demonstrates the runtime of the EF heuristic. On the largest instances ($n=20$, $r_{\max} = 16n$) it hits the one-hour limit. However, it doesn't disqualify the EF approach: we recall that the BDH runtime always equals 1 hour (3600 seconds), and the compact MILP is solved to optimality before the one-hour time limit only for $n=5$. 
    }
    \label{fig:obj_ef_vs_milp}
\end{figure*}

\begin{figure}[htbp]
\centerline{\includegraphics[width=0.5\textwidth]{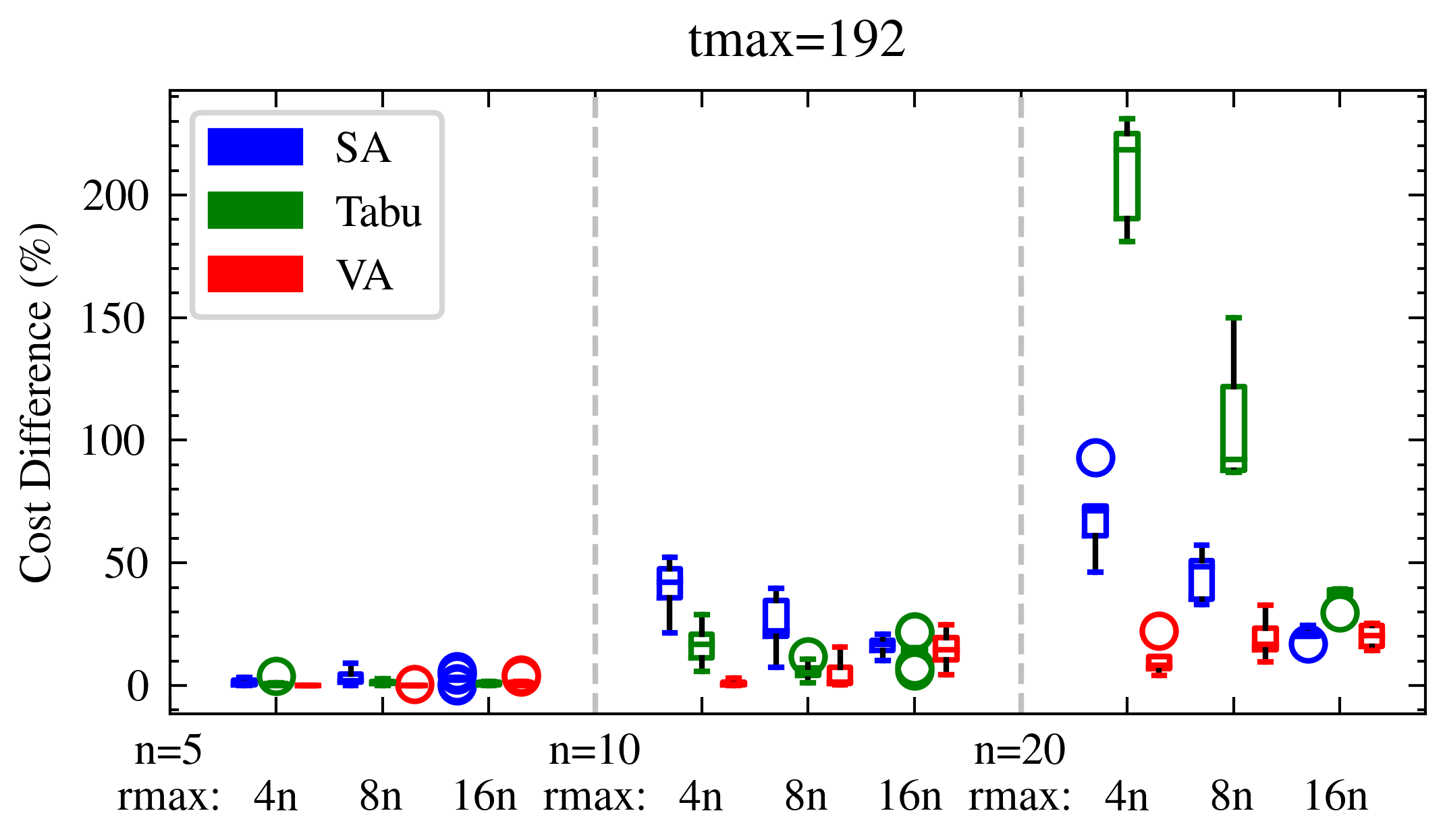}}
    \caption{Performance of classical metaheuristics on the generated instances of the Set Partition problem. (SA) stands for the simulated annealing, (Tabu)  for the tabu search, (VA) for the vector annealing by NEC.
    }
    \label{fig:sa_tabu_va}
\end{figure}

\subsection{Evaluation of the decomposed formulation}

On Figure \ref{fig:obj_ef_vs_milp} we compare the quality of solutions returned with the proposed decomposition scheme ("EF" for extended formulation) to the results found within one hour by the Benders-decomposition-based heuristic from \cite{Honda_paper}.
As a baseline, we take the solution obtained by Gurobi within one hour on the compact MILP formulation ($C^{MILP}$). 
In this experiment, we aim to evaluate the relevance of the decomposition (disregarding the performance of quantum solvers), so we delegate the master problem to the Gurobi solver (version 9.5). 
We report the relative difference in the cost value, $(C^h - C^{MILP}) / C^{MILP}$, where $C^{h}$ is the value of the objective function in the solution returned by heuristics $h:$ (BH) and (EF). 

We observe that on difficult instances the proposed EF approach leads to solutions of comparable quality while being significantly faster, except for the largest instances where the time needed for solution hits the one hour (3600 seconds) limits. We recall that on instances with $t_{\max} = 192$ the Benders-decomposition-based heuristic \textit{always} runs for \textit{one hour} disregarding the values of other parameters \cite{Honda_paper}.
Moreover, for large instances with $n=20$ vehicles the performance of EF approach scales better since the relative cost difference decreases with increasing $r_{\max}$. 
However, this is most likely due to the degrading performance of the reference MILP approach rather than an improved performance of the proposed heuristic approaches. 

\subsection{Metaheuristics for the master problem}

On large instances ($n=20$, $r_{\max} = 16n= 320$ ) the Gurobi solver reaches the 1-hour time limit while solving the integer master problem (Figure \ref{fig:obj_ef_vs_milp}, last panel). 
Therefore, we accelerate this step by moving from the exact Gurobi solver to the quantum annealing as well as to classical heuristic solvers.

We test the performance of the classical metaheuristics \textit{simulated annealing} \cite{SA_overview} and \textit{tabu search} \cite{Tabu} (Figure \ref{fig:sa_tabu_va})\footnote{We use the implementation from the dwave-neal module}. 
We also benchmark the quantum-inspired vector annealer by NEC\footnote{https://www.nec.com/en/global/quantum-computing/} on our instances of the Set Partition Problem.
The vector annealer performs the simulated annealing (on a vector supercomputer) but restricts local moves to the feasible subspace.
We observe that this modification significantly changes the behavior of the metaheuristic: while the standard simulated annealing on the QUBO formulation finds better solutions with an increasing number of reservations (and, as a consequence, the number of constraints in the master problem), the opposite is true for the vector annealer. 

Comparing the approximate solutions with the solutions found by Gurobi in one hour  (Fig.~\ref{fig:sa_tabu_va}), we observe that the  quality of the solution decreases substantially. The cost difference is always positive (no improvement) and for large instances is at least 10\% worse. 
However, given the substantially reduced runtime - each heuristic takes no more than 5 minutes - it might be reasonable to use heuristics and trade solutions quality for runtime improvement. 
As quantum annealers may further reduce the time-to-target \cite{DWave_benchmark}, in a close-to-online regime (where new reservations appear during the time-horizon) the hybrid approach is a promising option for cost-efficient planning.

\subsection{Quantum annealing for the master problem}

Finally, we evaluate the potential of our quantum-classical hybrid scheme on instances with $t_{\max} = 32$ where we use the DWave Advantage 6.1 for solving the Set Partition problem of the proposed decomposition scheme.
If the quantum annealer returns an infeasible solution we restore feasibility in a greedy fashion: subsets from the infeasible solution are iteratively added to a partial solution if the addition doesn't violate any constraints. 

We remark that even if we are able to run experiments only on the smallest instances from the benchmark dataset, we used the \textit{real data} and not the simplified one as in most papers \cite{Quantum_CG_master, tq_jsp} that benchmark the quantum annealer on industrial use-cases. 
We compared the obtained results to the ones found by QuEnc - the variational quantum algorithm for gate-based quantum computers based on \textit{amplitude encoding} \cite{quenc}. 

\begin{table}[hbtp]
\caption{Gap on $90$ instances for the Quantum Annealing (QA) and QuEnc}
    \centering
    \begin{tabularx}{0.5\textwidth}{|c|c|c|c|c|c|c|}
    \hline
  $n$ & $r_{\max}$  & QA Gap (\%) & \#physical$^{\mathrm{a}}$ & \#logical $^{\mathrm{a}}$ & QuEnc Gap (\%) \\
    \hline
1 & 4  & 0.00 & 4.70 & 4.70 & 0.00 \\
1 & 8  & 0.00 & 5.30 & 5.30 & 0.00 \\
1 & 16  & 0.00 & 5.80 & 5.60 & 0.00 \\
2 & 8  & 0.21 & 21.10 & 14.80 & 22.99 \\
2 & 16 & 0.00 & 15.90 & 13.20 & 6.31 \\
2 & 32  & 0.01 & 18.60 & 14.80 & 7.31\\
5 & 20  & \textbf{23.75} & 239.70 & 61.20 & \textbf{27.78} \\
5 & 40  & \textbf{9.59} & 251.00 & 80.80 & \textbf{9.50} \\
5 & 80  & \textbf{5.25} & 256.10 & 121.20 & \textbf{4.17} \\
    \hline 
    \multicolumn{6}{X}{
$^{\mathrm{a}}$ number of physical and logical qubits used by the quantum annealer.  
}
    \end{tabularx}
    \label{tab:gap_qa}
\end{table}

In the DWave Advantage hardware physical qubits interact only with their local neighbors in the \textit{Pegasus} layout. 
The QUBO has to be \textit{embedded} in the hardware architecture, which leads to an overhead since a \textit{logical qubit} $x_i$ has to be represented by a chain of \textit{physical qubits} $\{q_i^1, \dots, q_i^k\}$, see, e.g., \cite{yarkoni2022}. 

We observe (see Tab. \ref{tab:gap_qa}) that quantum annealing is able to find optimal or near-optimal solutions for very small instances of  $n=1$ and $n=2$ EVs. 
For $n=5$ the relative performance is worst for the smallest number of reservations $r_{\max}=20$ and improves with larger $r_{\max}$. 
Interestingly, this trend correlates with the embedding overhead.  
The QuEnc approach shows the same qualitative behavior as a function of $r_{\max}$ for fixed $n$, but already fails to find optimal solutions for $n=2$ EVs. The possible cause might be the generic hardware-efficient structure of QuEnc ansatz -- contrary to the problem-specific annealing evolution or cost-dependent QAOA  ansatz\cite{QAOA_original}. 

\section{Discussion}\label{section:discussion}

We suggested a new approach to address real-world problems with hybrid quantum-classical routines. Instead of formulating the problem as one MIP, we separate it into master and pricing problems; the NP-hard master problem is further delegated to a quantum (or hybrid) algorithm. Constraints that are difficult for quantum routines are managed inside the classical pricing routine. 

We tested our approach on the EVFCAP problem. 
The proposed decomposition of the original problem  into two sub-problems enables hybrid quantum-classical approaches despite the many inequality constraints in the compact formulation.
Additionally, for larger instances, it allowed us to find better solutions in a shorter time while using only classical methods.
Our numerical experiments also confirm that quantum annealing is in principle capable to solve the master problem. 
This spurs the hope that the integration of quantum routines can further accelerate the search for a good-quality approximate optimum in the future. 
However, experiments on hardware with more qubits and better connectivity are necessary to further evaluate the potential of a quantum advantage for this problem.



In this regard, the proposed approach provides a promising route to solve planning problems with difficult constraints with hybrid quantum-classical schemes.  

\bibliographystyle{IEEEtran}
\bibliography{References}

\end{document}